\documentclass[12pt,a4paper, amsmath,amssymb]{article}
\usepackage[margin=0.8in]{geometry}
\usepackage{longtable}
\usepackage{amssymb}
\usepackage{amsmath}
\usepackage{amsfonts}
\usepackage{bm}
\usepackage[utf8]{inputenc}
\usepackage{graphicx}
\usepackage{float}
\usepackage{xcolor}
\definecolor{bluemunsell}{rgb}{0.0, 0.5, 0.69}
\usepackage[colorlinks,linkcolor = bluemunsell,
urlcolor  = bluemunsell,
citecolor = bluemunsell,
anchorcolor = bluemunsell]{hyperref}
\usepackage{MnSymbol,bbding,pifont}
\usepackage{xcolor}
\usepackage{makecell}
\usepackage{ctable} 
\usepackage{adjustbox}
\def\thefootnote{\fnsymbol{footnote}}

\usepackage{cite}

\begin{document}
	\begin{center}
		{\Large \textbf{Explanation of the Hints for a 95 GeV Higgs Boson within a 2-Higgs Doublet Model\\ 
		}}
		\thispagestyle{empty} 
		\vspace{1cm}
		{\sc
			A. Belyaev$^{1,2}$\footnote{\href{mailto:a.belyaev@phys.soton.ac.uk}{a.belyaev@phys.soton.ac.uk}},
			R. Benbrik$^3$\footnote{\href{mailto:r.benbrik@uca.ac.ma}{r.benbrik@uca.ac.ma}},
			M. Boukidi$^3$\footnote{\href{mailto:mohammed.boukidi@ced.uca.ma}{mohammed.boukidi@ced.uca.ma}},
			M. Chakraborti$^1$\footnote{\href{mailto:mani.chakraborti@gmail.com}{mani.chakraborti@gmail.com}},
			S. Moretti$^{1,4}$\footnote{\href{mailto:s.moretti@soton.ac.uk}{s.moretti@soton.ac.uk}; \href{mailto:stefano.moretti@physics.uu.se}{stefano.moretti@physics.uu.se}},\\
			S. Semlali$^{1,2}$\footnote{\href{mailto:souad.semlali@soton.ac.uk}{souad.semlali@soton.ac.uk}}\\
		}
		\vspace{1cm}
		{\sl
			$^1$School of Physics and Astronomy, University of Southampton, Southampton, SO17 1BJ, United Kingdom.\\
			\vspace{0.1cm}	
			+		$^2$Particle Physics Department, Rutherford Appleton Laboratory, Chilton, Didcot, Oxon OX11 0QX, United Kingdom.\\
			\vspace{0.1cm}
			$^3$Polydisciplinary Faculty, Laboratory of Fundamental and Applied Physics, Cadi Ayyad University, Sidi Bouzid, B.P. 4162, Safi, Morocco.\\
			\vspace{0.1cm}
			$^4$Department of Physics and Astronomy, Uppsala University, Box 516, SE-751 20 Uppsala, Sweden.
		}
	\end{center}
	\vspace*{0.1cm}

	\begin{abstract}
	We suggest an explanation for and explore the consequences of the 
	excess  around 95 GeV in the   di-photon and di-tau invariant mass distributions recently reported  by the CMS collaboration at the Large Hadron Collider (LHC), together with the discrepancy that has long been observed at the Large Electron-Positron (LEP) collider in the $b\bar b$ invariant mass.  Interestingly, the most recent findings announced  by the ATLAS collaboration do not contradict, or even support, these intriguing observations. Their search in the di-photon final state similarly reveals an excess of events within the same mass range, albeit with a bit lower significance, thereby corroborating and somewhat reinforcing the observations made by CMS. 	
	We demonstrate that the lightest CP-even Higgs boson in the general 2-Higgs Doublet Model (2HDM) Type-III can explain simultaneously the observed excesses  at approximately 1.3 $\sigma$ C.L. while satisfying  up-to-date theoretical and experimental constraints.
	Moreover, the 2HDM Type-III  predicts an excess in the $pp\to t\bar t H_{\rm SM}$ production channel of the 125 GeV Higgs boson, $H_{\rm SM}$. This effect is caused by a up to 12\% enhancement of the $H_{\rm SM}tt$ Yukawa coupling in comparison to that predicted by the 
	Standard Model. Such an effect can be tested  at the High Luminosity LHC (HL-LHC), which can either discover or exclude the scenario we suggest. This unique characteristic of the 2HDM Type-III  makes this scenario with the  95 GeV resonance very attractive for further theoretical and experimental investigations at the (HL-)LHC and future  colliders.
		
	\end{abstract}

	\def\thefootnote{\arabic{footnote}}
	\setcounter{page}{0}
	\setcounter{footnote}{0}
	
	\newpage
	\section{Introduction}
\label{sec:intro}
In the last ten years, a great amount of effort has been put into the precise determination of the properties of the Higgs boson, following its discovery at the Large Hadron Collider (LHC)
in 2012~\cite{ATLAS:2012yve,CMS:2012qbp}. Many such properties are now established with accuracies better than 10\%  and most of the experimental observations made up to date are consistent with the Standard Model (SM) expectations. Nevertheless,  current precision Higgs physics at the LHC provides room for the possibility to go Beyond the SM (BSM) in search of additional Higgs states besides the SM-like 
one, with masses ranging from a few GeV up to the TeV scale.
Many well-motivated BSM scenarios with extended Higgs sectors, either  fundamental
(e.g., various Supersymmetric models \cite{Moretti:2019ulc}) or  effective ones (e.g., 2-Higgs Doublet Models (2HDMs)~\cite{Gunion:1992hs,Branco:2011iw}), predict the existence of extra light and heavy Higgs bosons, thus motivating  searches for these non-standard (pseudo)scalar states at various lepton and hadron colliders.

The 2HDM is one of the most well-studied BSM scenarios where the SM Higgs sector is extended by one additional Higgs doublet. In the most generic version of the 2HDM, the Yukawa couplings are non-diagonal in flavour space since each of the two Higgs doublets couple to all the SM fermions simultaneously. As a consequence, unwanted tree-level Flavour Changing Neutral Currents (FCNCs) may be induced, contradicting  experimental observations. To tackle this problem, usually a $Z_2$ symmetry is imposed on the model that determines in turn the coupling structure of the two Higgs doublets to the SM fermions, so that the 2HDM  can be classified into the so-called Type-I, Type-II, lepton-specific and flipped scenarios \cite{Branco:2011iw}. However, there is another possibility, i.e.,  the 2HDM Type-III where, instead of introducing such a $Z_2$ symmetry, one allows for  simultaneous couplings of the two Higgs doublets
to all  SM fermions. The generic Yukawa structure resulting from such a configuration can be constrained using various theoretical requirements of self-consistency of the model as well as a range of experimental measurements of the Higgs masses and couplings present in the model.

In the ongoing search for a low-mass Higgs boson, the CMS collaboration reported in 2018 an excess in the invariant mass of di-photon events   near 95 GeV \cite{CMS:2018cyk}. In March 2023, CMS has released their latest results, confirming the excess by employing advanced analysis techniques and 
utilising data collected during the first, second and third years of Run 2, corresponding to integrated luminosities of 36.3 fb$^{-1}$, 41.5 fb$^{-1}$ and 54.4 fb$^{-1}$, respectively, all at a center-of-mass energy of 13 TeV \cite{CMS:2023yay}. The combined data exhibited an excess with a local significance of 2.9$\sigma$ at a mass of $m_{\gamma\gamma}=95.4$ GeV.

The ATLAS collaboration recently released their findings in this channel  derived from the full Run 2 data set \cite{Arcangeletti}. Notably, their latest analysis showcases a significantly enhanced level of sensitivity compared to their previous study, which relied on only  80 fb$^{-1}$ of data \cite{ATLAS:2018xad}. In their updated analysis, ATLAS reveals an excess with a local significance of 1.7$\sigma$ in the $\gamma\gamma$ channel around an invariant mass of 95 GeV, remarkably aligning with the  reported CMS observation.

Furthermore, an additional
excess has  recently been reported by CMS in the search for a
light neutral (pseudo)scalar boson $\phi$ in the production and decay
process $gg,~b\bar b \to \phi \to \tau\tau $\cite{CMS:2022rbd}, with a
local (global) significance of $3.1\sigma (2.7\sigma)$ for $m_\phi\approx100~\text{GeV}$. Considering the poor resolution of $m_{\tau\tau}$ in comparison to $m_{\gamma\gamma}$,
the two excesses observed in the two final states actually appear
to be compatible. 
Previously, the Large Electron Positron (LEP) collider collaborations~\cite{LEPWorkingGroupforHiggsbosonsearches}
explored the low-mass domain extensively in the $e^+e^-\to Z\phi$ production mode, with a generic Higgs boson state $\phi$ decaying via the 
$\tau\tau$ and $b\overline{b}$ channels. Interestingly, an excess has been 
reported in {2006}
in the $e^+e^- \to Z {\phi}(\to b\overline{b})$ mode for  $m_{b\bar b}$ around 98 GeV~\cite{ALEPH:2006tnd}.
Given the limited mass resolution of the di-jet invariant mass at LEP, this anomaly may well coincide with the aforementioned excesses seen by CMS and/or ATLAS in the $\gamma\gamma$ and $\tau\tau$ final state. Since the excesses appear in very similar mass regions, several studies~\cite{Cao:2016uwt,Heinemeyer:2021msz,Biekotter:2021qbc,Biekotter:2019kde,Cao:2019ofo,Biekotter:2022abc,Iguro:2022dok,Li:2022etb,Cline:2019okt,Biekotter:2021ovi,Crivellin:2017upt,Cacciapaglia:2016tlr,Abdelalim:2020xfk,Biekotter:2022jyr,Biekotter:2023jld,Azevedo:2023zkg,Biekotter:2023oen} have explored the possibility of simultaneously explaining these anomalies within BSM frameworks featuring a non-standard Higgs state lighter than 125 GeV, while being in agreement with current measurements of the properties of the $\approx 125$ GeV SM-like Higgs state observed at the LHC. In the attempt to explain the excesses in the $\gamma\gamma$ and $b\overline{b}$ channels, it was found in Refs.~\cite{Benbrik:2022azi, Benbrik:2022tlg} that the 2HDM Type-III with a particular Yukawa texture can
successfully accommodate both measurements simultaneously with the lightest CP-even Higgs boson of the model, while being consistent with all relevant theoretical and experimental constraints. 
Further recent studies have shown that actually all three aforementioned signatures can be simultaneously explained in the 2HDM plus a real (N2HDM)~\cite{Biekotter:2022jyr} and complex (S2HDM) ~\cite{Biekotter:2023jld,Biekotter:2023oen} singlet.

%

{In our study, we show that all three anomalies seen in the $\gamma\gamma$, $\tau\tau$ and $b\bar b$ final states can also be  explained within the 2HDM Type-III of Refs.~\cite{Benbrik:2022azi, Benbrik:2022tlg}, 
	at approximately 1.3 $\sigma$ C.L.
	thereby making the point that one does not need to go beyond the minimal 2HDM framework.
	Moreover, in our study, we have found  an important prediction.
	The parameter space of 2HDM Type-III explaining the anomalous data,  predicts an enhancement of
	$H_{\rm SM}tt$ Yukawa coupling and the respective
	uplift of 
	the $gg,q\bar q\to t\bar t H_{\rm SM}$ production  of the SM-like Higgs, which can be tested in the near future to  discover or exclude the scenario we suggest.}

The paper is organised as follows. In section~\ref{sec:model}, we review the theoretical
framework we have chosen, i.e., the 2HDM Type-III. We describe the three excesses observed at  LEP and the LHC 
in section~\ref{sec:excess}. In section~\ref{sec:cons}, we outline the relevant theoretical
and experimental constraints considered in this analysis. In section~\ref{sec:results}, we present our numerical set-up to scan the parameter space of the 2HDM Type-III in order to find an explanation of the three anomalies and how this can be achieved, including the consequences for other Higgs processes.
Finally, we conclude in section~\ref{con}.
\section{2HDM Type-III}
\label{sec:model}
The 2HDM includes two $SU(2)_L$ doublets with hypercharge $Y=1$. The most general renormalisable $SU(2)_L\times U(1)_Y$ invariant scalar potential is written as follows~\cite{Branco:2011iw}:
\begin{eqnarray}
\mathcal{V} &= m_{11}^2 \Phi_1^\dagger \Phi_1+ m_{22}^2\Phi_2^\dagger\Phi_2 - \left[m_{12}^2
\Phi_1^\dagger \Phi_2 + \rm{H.c.}\right] + \lambda_1(\Phi_1^\dagger\Phi_1)^2 +
\lambda_2(\Phi_2^\dagger\Phi_2)^2 +
\lambda_3(\Phi_1^\dagger\Phi_1)(\Phi_2^\dagger\Phi_2)  ~\nonumber\\ &+
\lambda_4(\Phi_1^\dagger\Phi_2)(\Phi_2^\dagger\Phi_1) +
\frac12\left[\lambda_5(\Phi_1^\dagger\Phi_2)^2 +\rm{H.c.}\right] 
+\left\{\left[\lambda_6(\Phi_1^\dagger\Phi_1)+\lambda_7(\Phi_2^\dagger\Phi_2)\right]
(\Phi_1^\dagger\Phi_2)+\rm{H.c.}\right\}, \label{C2HDMpot}
\end{eqnarray}

where $m_{11}^{2}$, $m_{22}^{2}$, $m_{12}^{2}$ are mass squared terms
and $\lambda_{i}$ ($i = 1, ..., 7$) are dimensionless quantities describing the
coupling of the order-4 interactions. Of all such parameters, 6 are real ($m_{11}^{2}$,
$m_{22}^{2}$ and $\lambda_{i}$ with $i = 1, ..., 4$) and 4 are a priori complex
($m_{12}^{2}$ and $\lambda_{i}$ with $i = 5, ..., 7$). Therefore, in general,
the model has 14 free parameters. Under appropriate manipulations, this number
can however be reduced.
Following Ref.~\cite{Davidson:2005cw}, to start with, one can diagonalise the quadratic part of the potential
in the $({\Phi_1,\Phi_2})$ space, removing the $m_{12}^2$ term, thus getting
rid of 2 parameters. Then, one can make a relative $U(1)$
transformation on $\Phi_1$ or $\Phi_2$, making $\lambda_5$ real, hence down to 11 parameters.
Next, by removing  CP violation, the number of free parameters reduces to 9 (this requires making one neutral Higgs state 
decouple  from both $VV$ ($VV=W^+W^-$ and $ZZ$) and $H^+H^-$ interactions). 
Furthermore, the Yukawa matrices corresponding to the two doublets are not simultaneously
diagonalisable, which can pose a problem, as the off-diagonal elements lead to tree-level Higgs mediated Flavour Changing Neutral Currents (FCNCs) on which severe experimental bounds exist. The Glashow-Weinberg-Paschos (GWP) theorem~\cite{GWP1,GWP2} states that this type of FCNCs is absent  if at most one Higgs multiplet is responsible for providing mass to fermions of a given electric charge. This GWP condition can be enforced by a discrete 
${Z}_2$-symmetry ($\Phi_1 \rightarrow +\Phi_1$ and $\Phi_2\rightarrow -\Phi_2$) on the doublets, in which case the absence of FCNCs is natural. 
However, the need to allow for a 
softly broken ${Z}_2$-symmetry (in turn re-introducing a small $m^2_{12}$), as customarily done to enable
Electro-Weak Symmetry Breaking (EWSB) compliant with experimental measurements, relies on the existence of a basis where $\lambda_6 = \lambda_7$ = 0. As a consequence, altogether, one loses 2 parameters ($\lambda_6$ and $\lambda_7$) but regains 1 ($m^2_{12}$), thus 
reducing further their overall number down to 8.

Then, after EWSB has taken place, each scalar doublet acquires a Vacuum Expectation Value (VEV) that can be parametrised as follows:
\begin{equation}
\langle \Phi_{1} \rangle =
\frac{v}{\sqrt{2}}
\begin{pmatrix}  
0 \\
\cos\beta
\end{pmatrix}
\quad
\quad
\langle \Phi_{2} \rangle =
\frac{v}{\sqrt{2}}
\begin{pmatrix}  
0 \\
\sin\beta
\end{pmatrix},
\end{equation}
where the angle $\beta$ determines the ratio of the two doublet VEVs, $v_1$ and $v_2$, as  $\tan\beta = v_{2} / v_{1}$, and where
$v=246$ GeV is a fixed value, thereby giving a final count of 7 free parameters, which we choose to be: 
\begin{eqnarray}
m_h,~~m_H,~~m_A,~~m_{H^\pm},~~\sin(\beta - \alpha),~~\tan\beta,~~m^2_{12},
\label{eq:inputs}
\end{eqnarray}
where $\alpha$ is the mixing angle in the CP-even Higgs sector,  $m_H$ and $m_h$ denote the two CP-even Higgs  masses (with $m_h<m_H$ and where either of these can be the discovered SM-like Higgs state $H_{\rm SM}$\footnote{In our numerical analysis, we will assume $H\equiv H_{\rm SM}$.}) whereas $m_{H^\pm}$ and $m_A$ are the masses of the charged and CP-odd Higgs states, respectively. (In the remainder, we will use the short-hand notation $s_{x}$ and $c_x$ in place of $\sin(x)$ and $\cos(x)$.)

In the Yukawa sector, the general scalar to fermions couplings are given  by:
\begin{eqnarray}
-{\cal L}_Y &=& \bar Q_L Y^u_1 U_R \tilde \Phi_1 + \bar Q_L Y^{u}_2 U_R
\tilde \Phi_2  + \bar Q_L Y^d_1 D_R \Phi_1 
+ \bar Q_L Y^{d}_2 D_R \Phi_2 \nonumber \\
&+&  \bar L Y^\ell_1 \ell_R \Phi_1 + \bar L Y^{\ell}_2 \ell_R \Phi_2 + {\rm H.c.},
\label{eq:Yu}
\end{eqnarray}
where $Y^{f}_{1,2}$ are  $3\times 3$ Yukawa matrices in flavour space and $\tilde{\Phi}_{1,2}=i\sigma_{1,2}\Phi_{1,2}^*$, with $\sigma_{1,2}$ being the Pauli matrices. After EWSB has taken place, one can then derive the fermion masses from Eq.~(\ref{eq:Yu}). 

Here, however, we investigate a modified version of the described 2HDM, the so-called Type-III, where neither a global symmetry is implemented in the Yukawa sector nor any alignment in flavour space is enforced. We adopt instead the Cheng-Sher ansatz~\cite{Cheng:1987rs, Diaz-Cruz:2004wsi}, which assumes a flavour symmetry in turn suggesting a specific texture of the Yukawa matrices, where FCNC effects are proportional to the geometric mean of the two fermion masses and dimensionless  parameters\footnote{We refer the readers to Ref.~\cite{Hernandez-Sanchez:2012vxa} for more details.} $\chi_{ij}^{f}$ ($\propto \sqrt{m_i m_j}/ v ~\chi_{ij}^f$), 
where  $i,j=1-3$. After EWSB, the Yukawa sector  can be expressed in terms of the mass eigenstates of the Higgs bosons, as follows:
\begin{align}
-{\cal L}^{\rm III}_Y  &= \sum_{f=u,d,\ell} \frac{m^f_j }{v} \times\left( (\xi^f_h)_{ij}  \bar f_{Li}  f_{Rj}  h + (\xi^f_H)_{ij} \bar f_{Li}  f_{Rj} H - i (\xi^f_A)_{ij} \bar f_{Li}  f_{Rj} A \right)\nonumber\\  &+ \frac{\sqrt{2}}{v} \sum_{k=1}^3 \bar u_{i} \left[ \left( m^u_i  (\xi^{u*}_A)_{ki}  V_{kj} P_L+ V_{ik}  (\xi^d_A)_{kj}  m^d_j P_R \right) \right] d_{j}  H^+ \nonumber\\  &+ \frac{\sqrt{2}}{v}  \bar \nu_i  (\xi^\ell_A)_{ij} m^\ell_j P_R \ell_j H^+ + {\rm H.c.}\, \label{eq:Yukawa_CH},
\end{align}
where the $(\xi^f_{\phi})_{ij}$ couplings are given 
in Tab.~\ref{coupIII}  in terms of the free parameters $\chi_{ij}^f$,  $\tan\beta$ and the mixing angle $\alpha$. These expressions encompass the Higgs-fermion interactions of 2HDM Type-II
\footnote{The 2HDM Type-II is restored when $\chi_{ij}=0$.} together with  a contribution coming from the Yukawa texture\footnote{Here, $\xi_f^{\rm III} = \xi_f^{II} + \Delta_{ij}$, with $\Delta_{ij}\sim \chi_{ij}$ ~\cite{Hernandez-Sanchez:2012vxa,Crivellin:2013wna}.}. 
\begin{table*}[htb!]
	\begin{center}
		\setlength{\tabcolsep}{8pt}
		\renewcommand{\arraystretch}{0.8} %
		\begin{tabular}{c|c|c|c} \hline\hline 
			$\phi$  & $(\xi^u_{\phi})_{ij}$ &  $(\xi^d_{\phi})_{ij}$ &  $(\xi^\ell_{\phi})_{ij}$  \\   \hline
			$h$~ 
			& ~ $  \frac{c_\alpha}{s_\beta} \delta_{ij} -  \frac{c_{\beta-\alpha}}{\sqrt{2}s_\beta}  \sqrt{\frac{m^u_i}{m^u_j}} \chi^u_{ij}$~
			& ~ $ -\frac{s_\alpha}{c_\beta} \delta_{ij} +  \frac{c_{\beta-\alpha}}{\sqrt{2}c_\beta} \sqrt{\frac{m^d_i}{m^d_j}}\chi^d_{ij}$~
			& ~ $ -\frac{s_\alpha}{c_\beta} \delta_{ij} + \frac{c_{\beta-\alpha}}{\sqrt{2}c_\beta} \sqrt{\frac{m^\ell_i}{m^\ell_j}}  \chi^\ell_{ij}$ ~ \\
			$H$~
			& $ \frac{s_\alpha}{s_\beta} \delta_{ij} + \frac{s_{\beta-\alpha}}{\sqrt{2}s_\beta} \sqrt{\frac{m^u_i}{m^u_j}} \chi^u_{ij} $
			& $ \frac{c_\alpha}{c_\beta} \delta_{ij} - \frac{s_{\beta-\alpha}}{\sqrt{2}c_\beta} \sqrt{\frac{m^d_i}{m^d_j}}\chi^d_{ij} $ 
			& $ \frac{c_\alpha}{c_\beta} \delta_{ij} -  \frac{s_{\beta-\alpha}}{\sqrt{2}c_\beta} \sqrt{\frac{m^\ell_i}{m^\ell_j}}  \chi^\ell_{ij}$ \\
			$A$~  
			& $ \frac{1}{t_\beta} \delta_{ij}- \frac{1}{\sqrt{2}s_\beta} \sqrt{\frac{m^u_i}{m^u_j}} \chi^u_{ij} $  
			& $ t_\beta \delta_{ij} - \frac{1}{\sqrt{2}c_\beta} \sqrt{\frac{m^d_i}{m^d_j}}\chi^d_{ij}$  
			& $t_\beta \delta_{ij} -  \frac{1}{\sqrt{2}c_\beta} \sqrt{\frac{m^\ell_i}{m^\ell_j}}  \chi^\ell_{ij}$ \\ \hline \hline 
		\end{tabular}
	\end{center}
	\caption {Yukawa interactions in the 2HDM Type-III.  
	}  
	\label{coupIII}
\end{table*}

Based on the arguments presented above, the fundamental components of the Yukawa sector can be obtained in terms of the $\chi_{ij}^f$ parameters. These are additional free parameters of the model which describe masses and mixings of the  quarks and leptons. It is crucial to ensure that rare decays, which are suppressed in the SM, do not exceed current bounds, though. Specifically, it is important to investigate the contributions of Higgs bosons to FCNC processes in $B$ mesons. Here, the non-diagonal terms are not considered and the constraints from $\Delta B = 2$ processes can be ignored due to the suppression factor $\sqrt{m^f_j m^f_i}/v$. However, our analysis takes into account transitions involving $\Delta B = 1$ processes. The loop transition $b\to s\gamma$ is also sensitive to BSM physics, as deviations from the currently measured rate and SM predictions could indicate the presence of a light charged Higgs boson with appropriate Yukawa couplings. Finally, since $\rm{CP}$ violation is not observed in the lepton sector, it is reasonable to assume that the  $\chi_{ij}$'s are real numbers and the ensuing matrix is symmetric. However, 
the presence of these terms could also modify FCNCs in the Higgs sector, particularly $h\to f_i \bar{f_j}$ processes, which are  proportional to non-diagonal matrix terms \cite{Diaz-Cruz:2004wsi, Benbrik:2015evd}, so corresponding constraints need to be enforced. 

In the presence of the $\chi^f_{ij}$ texture parameters, alongside the standard 2HDM inputs of Eq.~(\ref{eq:inputs}), we will start our analysis by testing  the 2HDM Type-III against theoretical and current experimental constraints, which we do in the forthcoming section.

\section{The Excesses in $h \to \gamma\gamma,~\tau\tau$ and $b\bar{b}$ Channels}
\label{sec:excess}
In this section, we investigate whether the 2HDM Type-III can describe consistently the excess observed by both LEP and the LHC  at 94--100 GeV in the $\gamma\gamma$, $\tau\tau$ and $b\overline{b}$ channels. The evaluation of the so-called `signal strengths' for these excesses was done in the Narrow Width Approximation (NWA), in terms of the product of the relevant production cross section ($\sigma$, which at the LHC is dominated by gluon-gluon fusion through the top-quark loop\footnote{We have explicitly checked that in the region satisfying the flavour physics constraints, the allowed values of $\sin(\beta-\alpha)$ and $\chi_d^{33}$ lead to a reduction of $|c_{\phi b b}|$ when compared to $|c_{\phi t t}|$, so that the contribution from the bottom-quark loop is negligible (nearly $2\%$).} and at LEP by the Bjorken channel) as well as decay Branching Ratios (${\cal BR}$s) as follows:
\begin{eqnarray}
\mu_{{b\bar{b}}}&=&\frac{\sigma_{\rm 2HDM}(e^+e^-\to Z\phi )}{\sigma_{\rm SM}(e^+e^-\to Zh_{SM})}\times \frac{{\cal BR}_{\rm 2HDM}(\phi \to b\bar{b})}{{\cal BR}_{\rm SM}(h_{\rm SM}\to b\bar{b})} =\left|c_{\phi ZZ}\right|^2\times \frac{{\cal BR}_{\rm 2HDM}(\phi \to b\bar{b})}{{\cal BR}_{\rm SM}(h_{\rm SM}\to b\bar{b})},\label{mu_lep}\\\nonumber\\
\mu_{\mathrm{\tau\tau}}&=&\frac{\sigma_{\rm 2HDM}(gg\to \phi )}{\sigma_{\rm SM}(gg\to h_{\rm SM})}\times \frac{{\cal BR}_{\rm 2HDM}(\phi \to \tau\tau)}{{\cal BR}_{\rm SM}(h_{\rm SM}\to \tau\tau)} =\left|c_{\phi tt}\right|^2\times \frac{{\cal BR}_{\rm 2HDM}(\phi \to \tau\tau)}{{\cal BR}_{\rm SM}(h_{\rm SM}\to \tau\tau)},\\\nonumber\\
\mu_{\mathrm{\gamma\gamma}}&=&\frac{\sigma_{\rm 2HDM}(gg\to \phi )}{\sigma_{\rm SM}(gg\to h_{\rm SM})}\times \frac{{\cal BR}_{\rm 2HDM}(\phi \to \gamma\gamma)}{{\cal BR}_{\rm SM}(h_{\rm SM}\to \gamma\gamma)} =\left|c_{\phi tt}\right|^2\times \frac{{\cal BR}_{\rm 2HDM}(\phi \to \gamma\gamma)}{{\cal BR}_{\rm SM}(h_{\rm SM}\to \gamma\gamma)},\label{mu_cms}
\end{eqnarray} 	
where $c_{\phi ZZ}$ and $c_{\phi tt}$ are the $\phi$ couplings to $ZZ$ and $t\bar t$ (entering the LEP and LHC production modes, respectively) normalised to the corresponding SM values. In the present context, $\phi$ denotes the light CP-even Higgs scalar $h$ and our focus is exclusively on its contributions to the signal. We do not consider the pseudoscalar state $A$ for the explanation of these excesses in this analysis since for a CP-conserving scenario such as ours, the $AZZ$ coupling is forbidden at tree level, rendering the explanation of the LEP excess with a CP-odd state impossible. The experimental measurements for the three signal strengths are expressed as \cite{Biekotter:2023oen,Biekotter:2022jyr,Biekotter:2023jld}\footnote{The value of $\mu_{\gamma\gamma}^{\mathrm{exp}}$ 
	which we have decided to take form the latest version of~\cite{Biekotter:2023oen} approximately agrees with our rough estimation which we have found to be about 0.3.}:

\begin{eqnarray}
\mu_{\gamma\gamma}^{\mathrm{exp}}=\mu_{\gamma\gamma}^{\mathrm{ATLAS+CMS}} = 0.24^{+0.09}_{-0.08},
\quad \quad\quad \mu_{\tau\tau}^{\mathrm{exp}} = 1.2 \pm 0.5, \quad \quad\quad  \mu_{b\bar{b}}^{\mathrm{exp}} = 0.117 \pm 0.057,
\end{eqnarray}
where $h_{\rm SM}$ corresponds to a SM-like Higgs boson with a mass of $95$ GeV -- the mass  of the $h$ state of our interest from the 2HDM Type-III.
In our analysis, we have combined the di-photon measurements from the ATLAS and CMS experiments, denoted as $\mu_{\gamma\gamma}^{\mathrm{ATLAS}}$ and $\mu_{\gamma\gamma}^{\mathrm{CMS}}$, respectively. The ATLAS measurement yields a central value of $0.18{\pm0.1}$ \cite{Biekotter:2023oen} while the CMS measurement yields a central value of $0.33^{+0.19}_{-0.12}$\cite{Biekotter:2023jld}. By doing so, we aimed at leveraging the strengths of both experiments and improve the precision of our analysis. The combined measurement, denoted as $\mu_{\gamma\gamma}^{\mathrm{ATLAS+CMS}}$, is determined by taking the average of the central values without assuming any correlation between them. To evaluate  the combined uncertainty we sum ATLAS and CMS uncertainties in quadrature.

To determine whether a simultaneous fit to the observed excesses is possible, a $\chi^2$ analysis is performed using measured central values $\mu^{\mathrm{exp}}$ and the 1$\sigma$ uncertainties $\Delta\mu^{\mathrm{exp}}$ of the signal rates related to the two excesses as defined in Eqs. (\ref{mu_lep})--(\ref{mu_cms}). The contribution to the $\chi^2$ value
for each  channel ($\gamma\gamma$, $\tau\tau$, $b\bar{b}$) is calculated using the formula
\begin{eqnarray}
\chi^2_{\gamma\gamma,\tau\tau,b\bar{b}}=\frac{\left(\mu_{\gamma\gamma,\tau\tau,b\bar{b}}-\mu_{\gamma\gamma,\tau\tau,b\bar{b}}^\mathrm{ exp}\right)^2}{\left(\Delta\mu^\mathrm{exp}_{\gamma\gamma,\tau\tau,b\bar{b}}\right)^2}.
\end{eqnarray}
{So, the resulting $\chi^2$
	which we will use to judge whether the points from the model describe the 
	excess in  three channels, reads as:
}
\begin{equation}
\chi^2_{\gamma\gamma+\tau\tau+b\bar{b}}=
\chi^2_{\gamma\gamma}+\chi^2_{\tau\tau}+\chi^2_{b\bar{b}}.
\label{eq:chi95}
\end{equation}

\section{Theoretical and Experimental Constraints}
\label{sec:cons}

In our study we use a comprehensive set of theoretical and experimental constraints that must be satisfied to establish a viable model. 

\subsection{Theoretical Constraints} 
In our study we impose the following set of the theoretical constraints on the scalar potential.
\begin{itemize}
	\item \textbf{Unitarity}: The scattering processes involving (pseudo)scalar-(pseudo)scalar, gauge-gauge and/or (pseudo)scalar-gauge initial and/or final states must satisfy unitarity constraints. The eigenvalues $e_i$ of the tree-level 2-to-2 body scattering matrix should meet the following criteria: $|e_i|<8\pi$ ~\cite{uni1,uni2}.
	
	\item \textbf{Perturbativity}: Adherence to perturbativity constraints imposes an upper limit on the quartic couplings of the Higgs potential: $|\lambda_i|<8\pi$ ~\cite{Branco:2011iw}.
	
	\item \textbf{Vacuum Stability}: The scalar potential must be positive and bounded from below in any direction of the fields $\Phi_i$ to ensure vacuum stability. This requires that $\lambda_1>0$, $\lambda_2>0$, $\lambda_3>-\sqrt{\lambda_1\lambda_2}$, and $\lambda_3+\lambda_4-|\lambda_5|>-\sqrt{\lambda_1\lambda_2}$ ~\cite{Barroso:2013awa,sta}.
\end{itemize}

\subsection{Experimental Constraints} 
We also apply a variety of experimental constraints from EW Precision Observables (EWPOs), measurements of the observed Higgs boson properties at the LHC, lack of signals from non-SM-like Higgs bosons at LEP, Tevatrron and LHC as well as flavour observables,
which include the following.
\begin{itemize}
	\item \textbf{EWPOs}: We require a $95\%$ C.L. in matching the global fit results of the EW oblique parameters $S, T$ and $U$ \cite{Grimus:2007if,oblique2} with the following values~\cite{particle2020review}:
	\begin{align}
	S = 0.05 \pm 0.08,\quad T = 0.09 \pm 0.07,\nonumber\quad \rho_{ST}=0.92~~(\mathrm{for}~U=0).\hspace{1.5cm} 
	\end{align}
	
	\item\textbf{SM-like Higgs Boson Discovery}: 
	
	We made sure that the points from our parameter space agree with the experimental measurement of the properties of the discovered SM-like Higgs boson with mass of $\approx 125$ GeV red at 95\% C.L.
	To do this we have used the
	public code \texttt{HiggsSignals-3} \cite{Bechtle:2020pkv,Bechtle:2020uwn} via \texttt{HiggsTools} \cite{Bahl:2022igd} to perform a $\chi^2$
	test in order to check  how the Higgs signal strengths  from Tevatron and  LHC are compatible with the model predictions.

	\item \textbf{Non-SM-like Higgs Boson Exclusions}: To further scrutinise the parameter space of our 2HDM Type-III, we subject the selected parameter space points to rigorous examinations against exclusion limits derived from additional Higgs boson searches. We utilise the  code \texttt{HiggsBounds-6} \cite{Bechtle:2008jh,Bechtle:2011sb,Bechtle:2013wla,Bechtle:2015pma} via \texttt{HiggsTools} to incorporate the exclusion constraints from various experiments, including LEP, Tevatron and the LHC. 
	
	\item\textbf{$B$-Physics Observables}: We test various  $B$-physics observables against experimental data using the public code \texttt{SuperIso\_v4.1} \cite{superIso}. The following experimental measurements are used in our analysis:
	
	\begin{table}[H]
		\begin{center}
			\setlength{\tabcolsep}{22pt}
			\renewcommand{\arraystretch}{1.5}
			\begin{adjustbox}{max width=0.86\textwidth}		
				\begin{tabular}{lcc}
					\hline\hline
					Observable &   Value & Reference\\\hline\hline
					${\cal BR}(\overline{B}\to X_s\gamma)|_{E_\gamma<1.6\mathrm{~GeV}}$&$\left(3.32\pm0.15\right)\times 10^{-4}$&\cite{HFLAV:2016hnz}\\
					${\cal BR}(B^+\to \tau^+\nu_\tau)$&$\left(1.06\pm0.19\right)\times 10^{-4}$&\cite{HFLAV:2016hnz}\\
					${\cal BR}(D_s\to \tau\nu_\tau)$&$\left(5.51\pm0.18\right)\times 10^{-2}$&\cite{HFLAV:2016hnz}\\
					${\cal BR}(B_s\to \mu^+\mu^-)$ (LHCb)&$\left(3.09^{+0.46}_{-0.43}\right)\times 10^{-9}$&\cite{LHCb:2021awg,LHCb:2021vsc}\\
					${\cal BR}(B_s\to \mu^+\mu^-)$ (CMS)&$\left(3.83^{+0.38}_{-0.36}\right)\times 10^{-9}$&\cite{CMS:2022mgd}\\
					${\cal BR}(B^0\to \mu^+\mu^-)$ (LHCb)&$\left(1.2^{+0.8}_{-0.7}\right)\times 10^{-10}$&\cite{LHCb:2021awg,LHCb:2021vsc}\\
					${\cal BR}(B^0\to \mu^+\mu^-)$ (CMS)&$\left(0.37^{+0.75}_{-0.67}\right)\times 10^{-10}$&\cite{CMS:2022mgd}\\
					${\cal BR}(K\to \mu\nu_\mu)/{\cal BR}(\pi\to \mu\nu_\mu)$&$0.6357 \pm 0.0011$ &\cite{LHCb:2017rmj}\\\hline\hline
				\end{tabular}
			\end{adjustbox}
			\caption{Flavour physics observables and corresponding values employed in our analysis.}
		\end{center}
	\end{table}
	By rigorously examining these $B$-physics observables against experimental constraints, we can validate the compatibility of our 2HDM Type-III with  existing data and potentially uncover any deviations that could indicate new physics phenomena.
\end{itemize}

\section{Explanation of the Excesses}\label{sec:results}
In this section, we present  our numerical analysis of the 
2HDM Type-III parameter space. For the 2HDM Type-III spectrum generation, we have employed \texttt{2HDMC 1.8.0}\cite{2HDMC}, which considers the theoretical constraints discussed in the previous section, along with the electroweak precision observables (EWPOs). Subsequently, we validate our results by comparing them to Higgs data, utilizing \texttt{HiggsTools} \cite{Bahl:2022igd}, which includes the most recent versions of both \texttt{HiggsBounds} and \texttt{HiggsSignals}.
In accordance with the above discussions,
we consider the scenario where the heavier CP-even Higgs boson $H$ is the SM-like Higgs
particle $H_{\rm SM}$ discovered at the LHC with $m_{H_{\rm SM}}\approx$  125 GeV. In this scenario the lighter CP-even Higgs, $h$, is the source of the observed excess
in $\gamma\gamma$, $\tau\tau$ and $b\bar{b}$ channels around 95 GeV, which we previously labelled as $h_{\rm SM}$.
To explore this scenario, we conducted a systematic random scan across the parameter ranges specified in Tab.~\ref{tab:par-scan}\footnote{In this work, we do not include the contributions to the signal from the CP-odd state.}.

\begin{table}[H]
	\centering
	\begin{adjustbox}{max width=\textwidth}
		\begin{tabular}{c|c|c|c|c|c|c|c}\hline\hline
			$m_h$&$m_H$&$m_A$&	$m_{H^\pm}$& $s_{\beta-\alpha}$&$\tan\beta$&$m_{12}^2 $&$\chi_{ij}^{f,\ell} $ \\\hline
			$[94;\,97]$&$125.09$&$[80;\,300]$&$[160;\,200]$& $[-0.5;\,0]$&$[1;\,30]$&$m_h^2\tan\beta/(1+{\tan^{2}\beta})$&$[-3;\,3]$\\\hline\hline
		\end{tabular}
	\end{adjustbox}	
	\caption{Scan ranges of the 2HDM Type-III input parameters. Masses are given in GeV.} \label{tab:par-scan}
\end{table}
\noindent
We then investigate parameter spaces that satisfy the condition $\chi^2_{125}\le 189.4$, corresponding to a 95\% Confidence Level (C.L.) for 159 degrees of freedom, where $\chi^2_{125}$ corresponds to the $\chi^2$ evaluated by \texttt{HiggsSignals} for the 125 GeV Higgs signal strength measurements. Subsequently, we examine 2-dimensional (2D) planes of the signal strength parameters: $(\mu_{\gamma\gamma}-\mu_{\tau\tau})$, ($\mu_{\gamma\gamma}-\mu_{b\bar b}$) 
and ($\mu_{b\bar b}-\mu_{\tau\tau}$).

\begin{figure}[H]
	\centering
	\includegraphics[width=\textwidth]{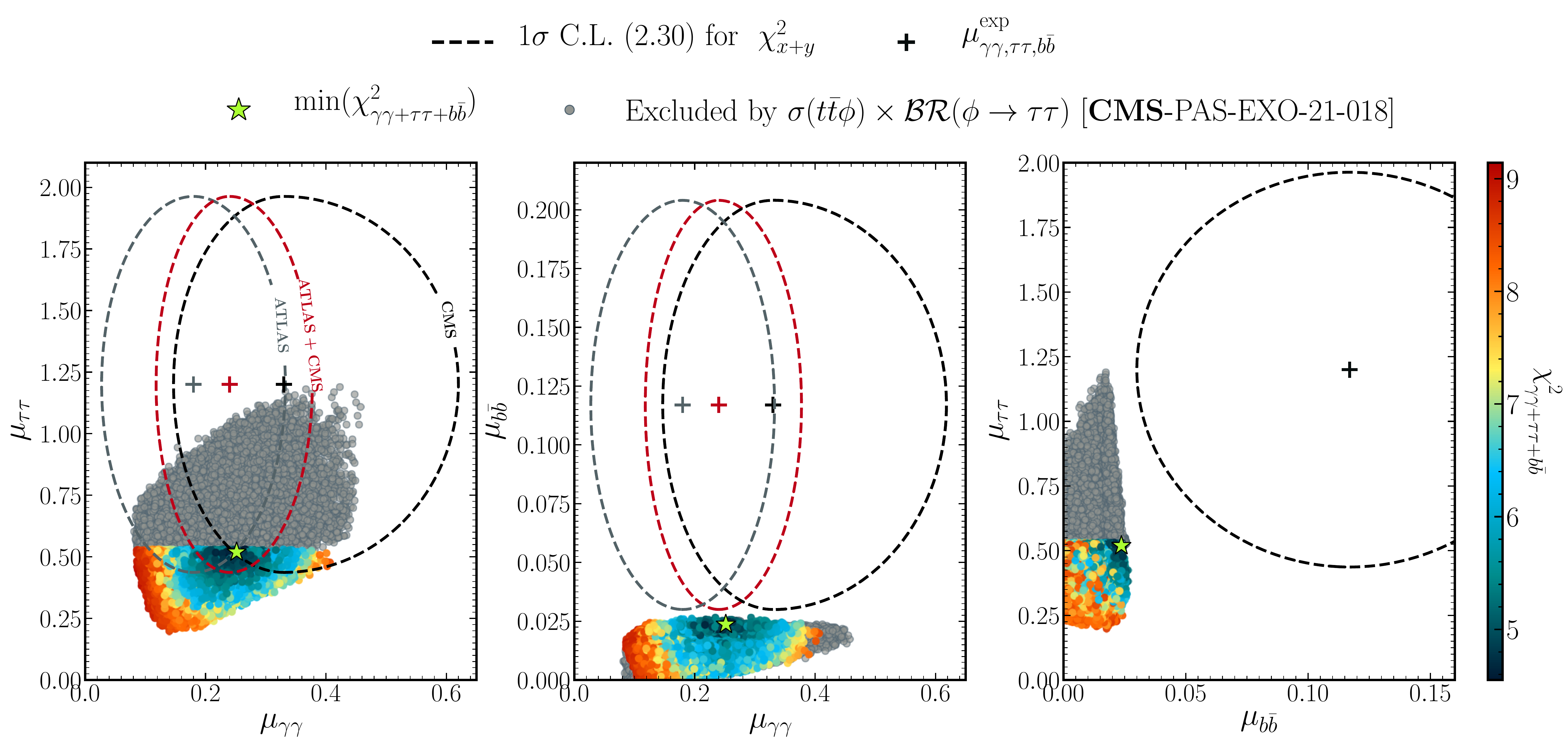}	
	\caption{The colour map of $\chi^2_{\gamma\gamma+\tau\tau+b\bar{b}}$ in the 
		($\mu_{\gamma\gamma}-\mu_{\tau\tau})$, ($\mu_{\gamma\gamma}-\mu_{b\bar b}$) 
		and ($\mu_{b\bar b}-\mu_{\tau\tau}$)
		planes of the 
		signal strength parameters for  2HDM Type-III parameter space under study.
		The  ellipses define the
		regions consistent with the excess at 
		1$\sigma$ C.L.
		The contribution from $\mu_{\gamma\gamma}$ to $\chi^2$
		for  black, gray and red contours comes from CMS, ATLAS and combined CMS+ATLAS data respectively.
		As detailed in the text, the contribution to $\chi^2$ from $\mu_{\tau\tau}$ (left frame)
		comes from CMS data while the contribution from 
		$\mu_{b\bar{b}}$  (middle frame) comes from LEP data. {Grey points are excluded by $\sigma(tt\phi) \times \mathcal{BR}(\phi \to \tau\tau)$ \cite{CMS:2022arx}}. The position of  $\chi^2_{95\mathrm{min}}$  is marked by a green star.}
	\label{fig1}
\end{figure}
In the following we will refer to the $h$ and $H$ states of the 2HDM Type-III by using the labels `$h_{95}$' and `$h_{125}$', respectively. In Fig.~\ref{fig1}, we present the results for 
$\chi^2_{\gamma\gamma+\tau\tau+b\bar{b}}$
in the form of its colour map projected into  the 
($\mu_{\tau\tau}$-$\mu_{\gamma\gamma}$) (left),
($\mu_{b \bar b}$-$\mu_{\gamma\gamma}$) (middle) and
($\mu_{\tau\tau}$-$\mu_{b \bar b}$) (right)  planes of the 
signal strength parameters.
The dashed ellipses define the
regions consistent with the excess at 
1$\sigma$  
described 
by the $\chi^2_x+\chi^2_y=2.30$ 
equation, where the subscripts $x$ and $y$
label each possible pairing out of  three signal channels ($\gamma\gamma$, $\tau\tau$ and $b\bar{b}$), depending on the frame of Fig.~\ref{fig1}.
The black, gray and red contours are for the $\chi^2$ constructed
using the $\mu_{\gamma\gamma}^{\mathrm{CMS}}$, $\mu_{\gamma\gamma}^{\mathrm{ATLAS}}$ and  $\mu_{\gamma\gamma}^{\mathrm{CMS+ATLAS}}$ signal strengths, respectively. 
The value of $\chi^2_{\gamma\gamma+\tau\tau+b\bar{b}}$ is indicated by the vertical colourmap.
The grey points represent exclusions based on recent CMS searches\footnote{This limit has not yet been integrated into \texttt{HiggsBounds-6}.} \cite{CMS:2022arx} for the production of a Higgs boson in association with either a top-quark pair or a $Z$ boson, with the subsequent decay into a tau pair. {The green star, indicating the position of $\chi^2_{95, \mathrm{min}}$ which is the minimum of $\chi^2_{\gamma\gamma+\tau\tau+b\bar{b}}$, has a value of 4.55, corresponds to a 1.26 $\sigma$ C.L. for three degrees of freedom}  even though this minimum lies solely within the 1$\sigma$ C.L. contour for the $\mu_{\tau\tau}$-$\mu_{\gamma\gamma}$ pair of signal strengths and not for the other pairs.
Furthermore, numerous points surrounding $\chi^2_{95, \mathrm{min}}$ depicted by the dark blue colour indicating the capability of the 2HDM Type-III to explain the observed excess across all three channels simultaneously at {$1.5\sigma$ C.L or better reaching up to about 1.3$\sigma$}.

One can also note from the middle and the right frames of Fig.~\ref{fig1}
that  all points are situated outside of the 1$\sigma$ ellipses because
it is difficult to achieve large enough values of $\mu_{b\bar{b}}$ satisfying the
three constraints simultaneously.
This happens because the $\gamma\gamma$ excess is achieved 
via enhancement of  ${\cal BR}(h_{95}\to \gamma\gamma)$
due to the decrease of $\Gamma_{b\bar b}$ (the main channel of $h_{95}$ 
decay) and the respective decrease of 
${\cal BR}(h_{95}\to b\bar b)$.

\begin{figure}[H]
	\begin{minipage}{0.45\textwidth}
		\centering		\includegraphics[height=8.cm,width=7.25cm]{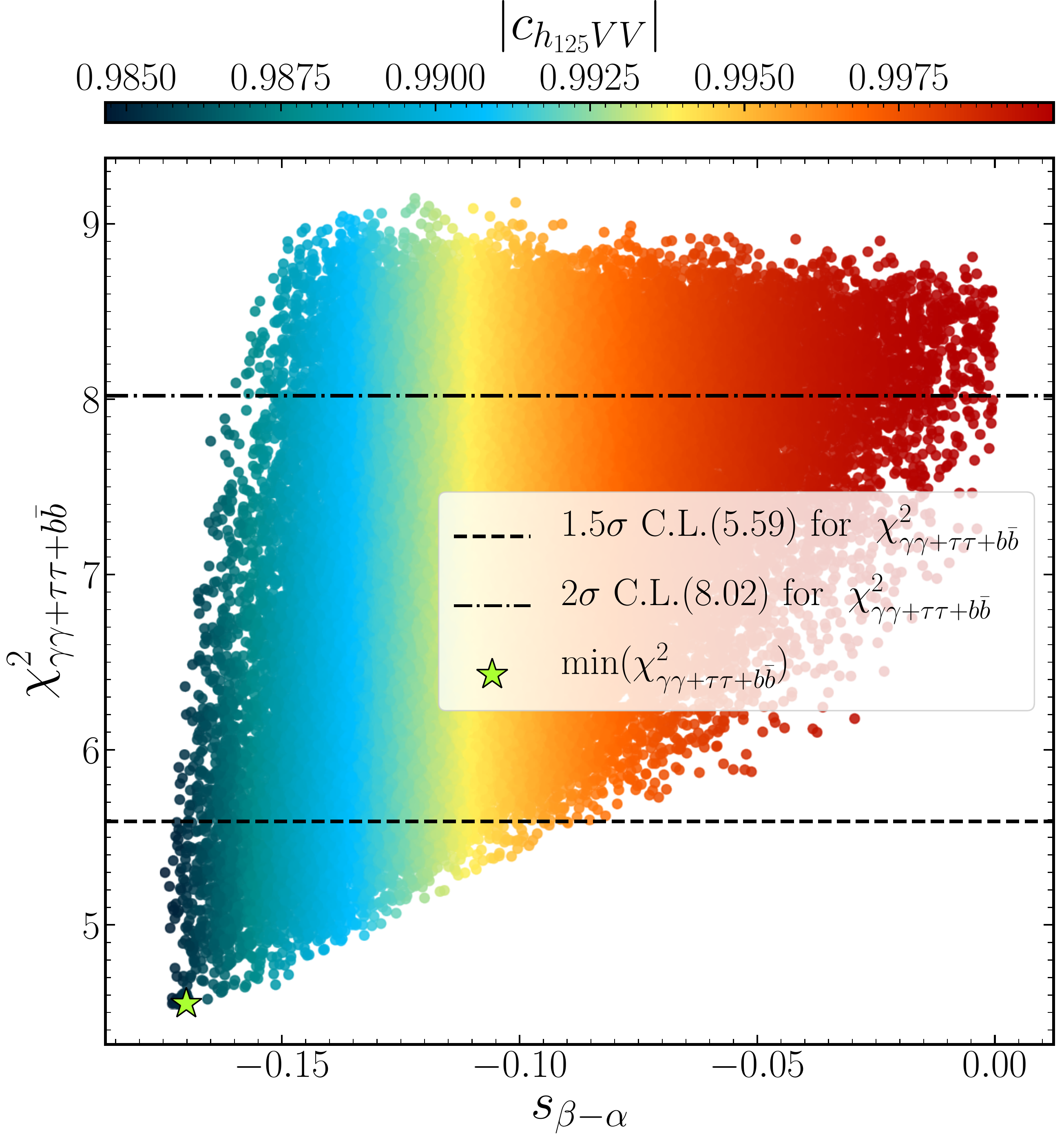}
	\end{minipage}\hspace{0.75cm}
	\begin{minipage}{0.45\textwidth}
		\centering		\includegraphics[height=8.cm,width=7.25cm]{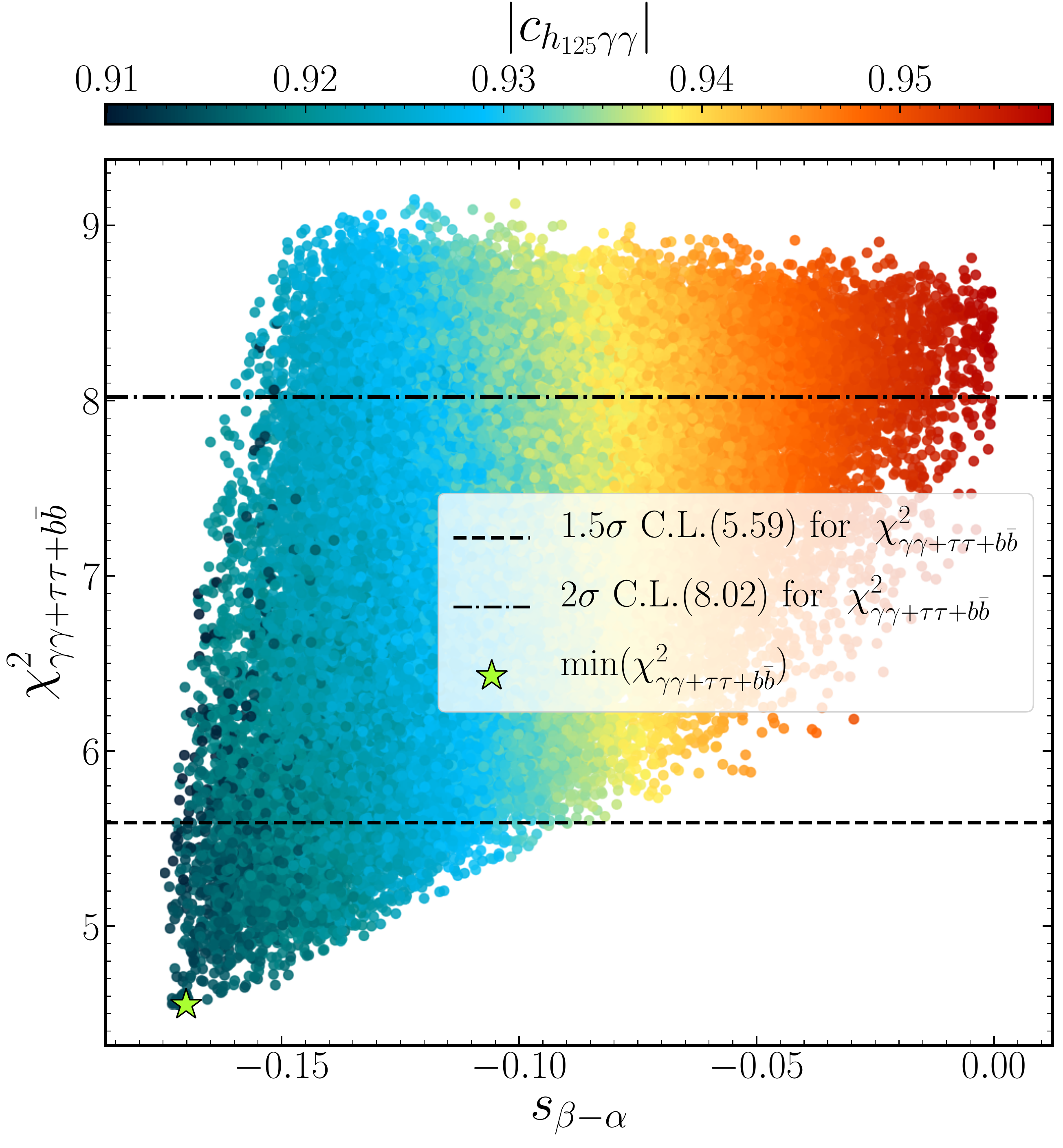}
	\end{minipage}			
	\caption{The $\chi^2_{\gamma\gamma+\tau\tau+b\bar{b}}$  dependence upon $s_{\beta-\alpha}$.
		The colour bar indicates  the value of  $|c_{h_{125}VV}|$ (left) and  $|c_{h_{125}\gamma\gamma}|$ (right). The horizontal dashed(dash-dot) line represents
		the 1.5$\sigma$(2$\sigma$) region.}\label{fig2}
\end{figure}

In Fig.~\ref{fig2}, we show $\chi^2_{\gamma\gamma+\tau\tau+b\bar{b}}$ in our
parameter scans as a function of $s_{\beta-\alpha}$.
We also indicate the value of the Higgs couplings to SM gauge bosons,
$|c_{h_{125}VV}|$ (left) and  $|c_{h_{125} \gamma\gamma}|$\footnote{The normalised coupling for the loop-induced channel ${h\to \gamma\gamma}$ is defined by $|c_{h\to \gamma\gamma}|^2\equiv\frac{\Gamma(h\to\gamma\gamma)^{\mathrm{2HDM}}}{\Gamma(h\to\gamma\gamma)^{\mathrm{SM}}}$.}
(right) in the colour bar.
The horizontal dashed(dash-dot) line represents
the 1$\sigma$(2$\sigma$) region corresponding to the three excesses ($\gamma\gamma,~\tau\tau,~b\bar{b}$).
Clearly one can read from the left panel that, when simultaneously describing the three excesses  {at 1.3$\sigma$ C.L.}, the SM-like Higgs coupling to vector bosons $|c_{h_{125}VV}|$ which is proportional to $c_{\beta-\alpha}$ lie close  to $\sim 1$, In contract the coupling to $\gamma\gamma$ exhibits a slight deviation from the predicted SM value, reaching a minimum of {0.91} and a maximum of 0.96.

\begin{figure}[H]
	\begin{minipage}{0.45\textwidth}
		\centering
		\includegraphics[height=8.cm,width=7.25cm]{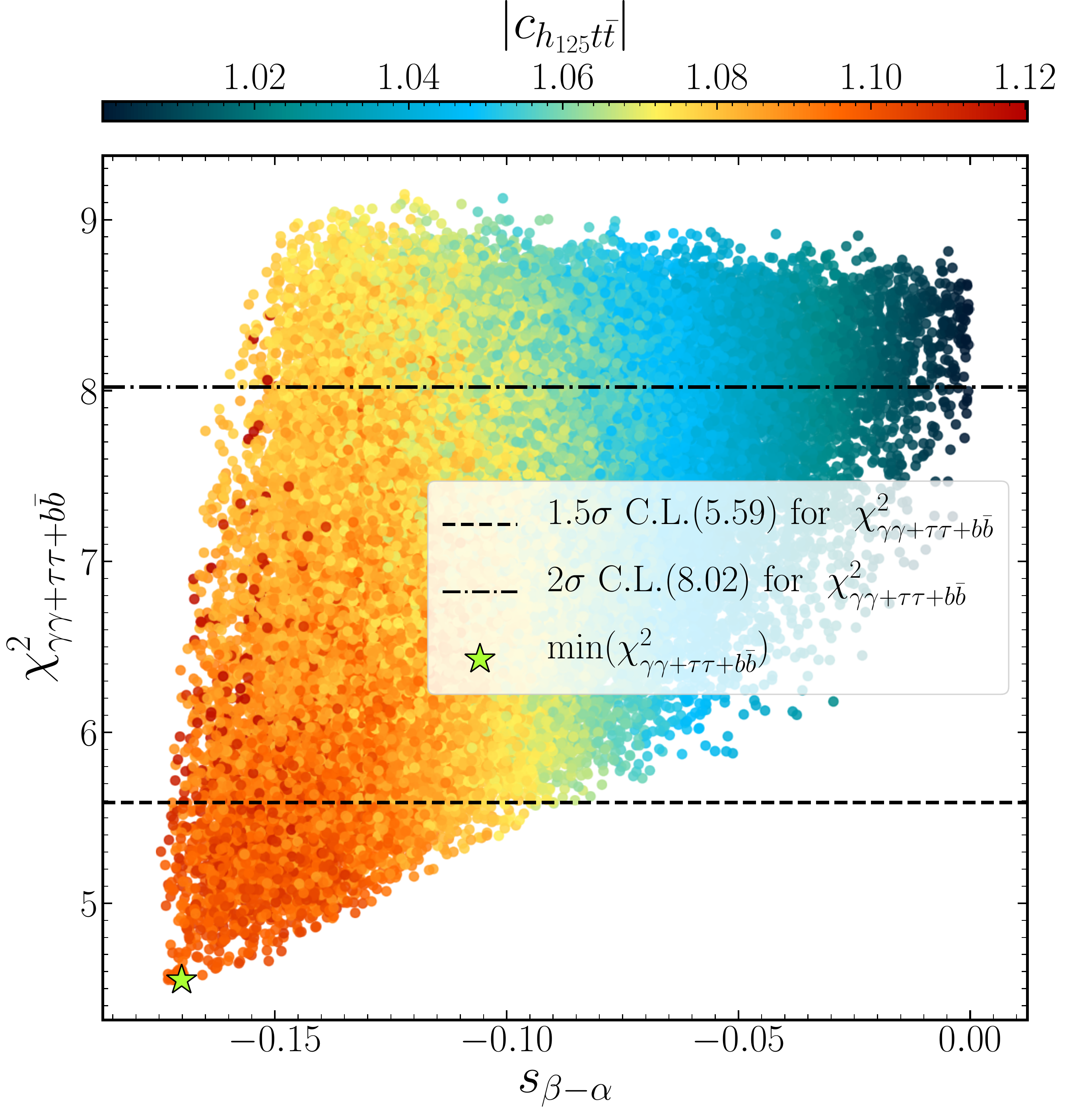}
	\end{minipage}\hspace{0.75cm}
	\begin{minipage}{0.45\textwidth}
		\centering
		\includegraphics[height=8.cm,width=7.25cm]{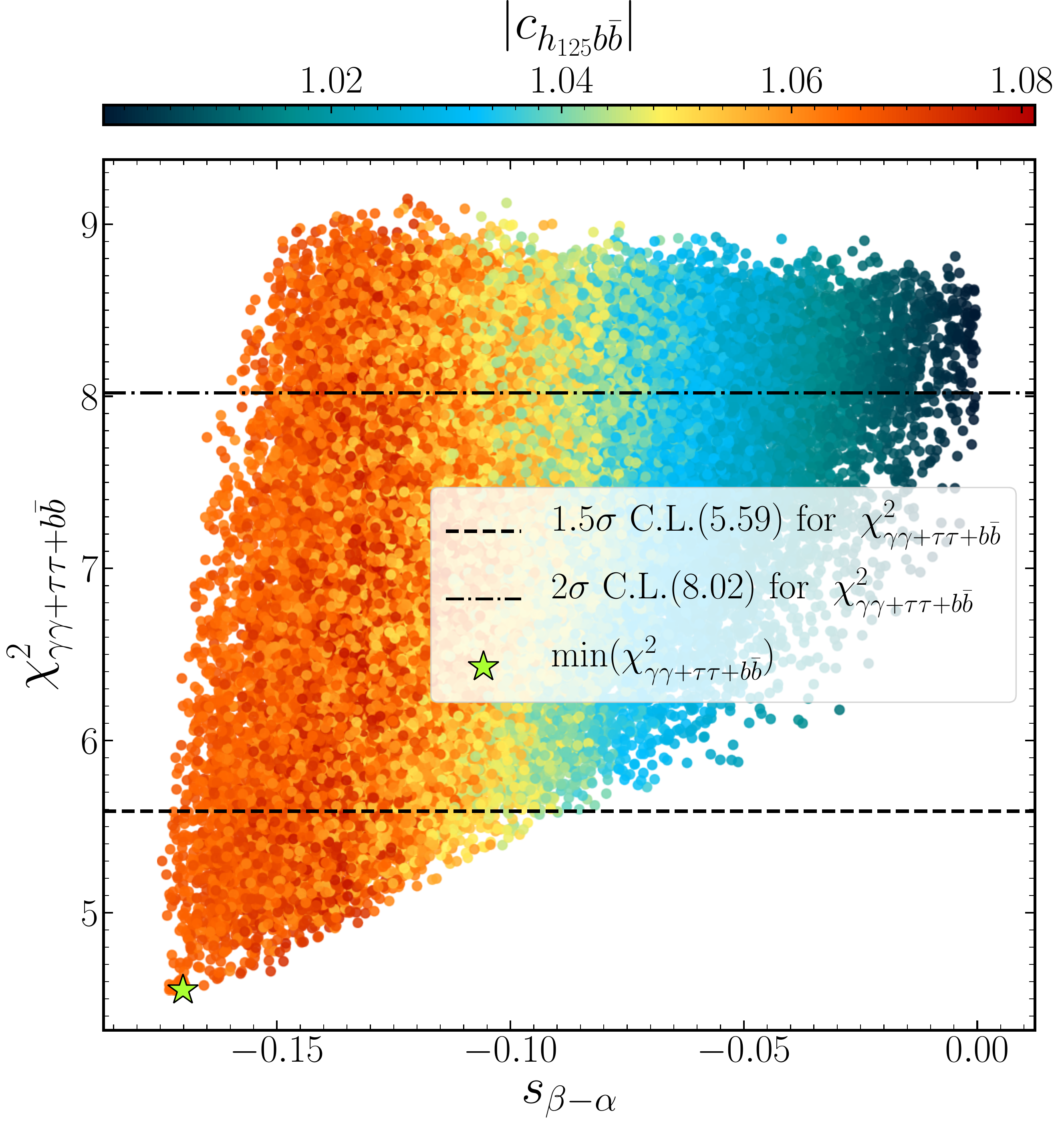}
	\end{minipage}			
	\caption{As in Fig.~\ref{fig2}, but the colour coding indicates the
		value of $|c_{h_{125}t\bar{t}}|$ (left) and  $|c_{h_{125}b\bar{b}}|$ (right).}\label{fig3}
\end{figure}

Fig.~\ref{fig3} presents analogous plots to those displayed in Fig.~\ref{fig2}. However, in this case, the colour bar represents the values of the couplings of the $\approx 125$ GeV
Higgs state to $t\bar{t}$ (left) and to $b\bar{b}$ (right).  The explanation of the three excesses simultaneously requires an enhancement of up to 12\% and 8\% in $c_{h_{125}t\bar{t}}$ and $c_{h_{125}b\bar{b}}$, respectively. The reason for the enhancement in $c_{h_{125}t\bar{t}}$ can be understood as follows. The requirement of fitting the three excesses at 1.3$\sigma$ C.L.
restricts $\sin(\beta-\alpha)$ in the range $\approx$ [-0.18,-0.13]. This makes the ${s_\alpha}/{s_\beta}$ term in the $c_{h_{125}t\bar{t}}$ coupling (see second row, Tab.~\ref{coupIII}) slightly larger than one. The $\chi_{33}^u$ parameter in the second term is constrained by flavour physics to be both small and negative, making the contribution from the second term minimal. In contrast, the enhancement in $c_{h_{125}b\bar{b}}$ originates from both terms, as $\chi_{33}^d$ is larger, contributing substantially to the coupling. A typical set of values for the $\chi_{ij}^f$ parameters can be found in Tab. \ref{Bp} where we describe the features of our best fit point. The implications of the enhancements, in particular in the $c_{h_{125}t\bar{t}}$ coupling, will be discussed  subsequently.

\noindent
In Fig.~\ref{fig5}, we directly compare our allowed parameter points to the
experimental data by superimposing these onto the CMS 13 TeV low-mass $\gamma\gamma$ \cite{CMS:2023yay}
analysis data.
{The light green colour represents the parameter points that fit the excesses within a three-dimensional C.L. {from $1.5\sigma$ up to $1.3\sigma$}, whereas the points that fit the excesses at
	2$\sigma$ or more are shown in red. 
	It can be clearly observed from the plots that our parameter points
	are exactly  suited to satisfy the excesses.
	\begin{figure}[h!]
		\centering
		\includegraphics[width=0.95\textwidth]{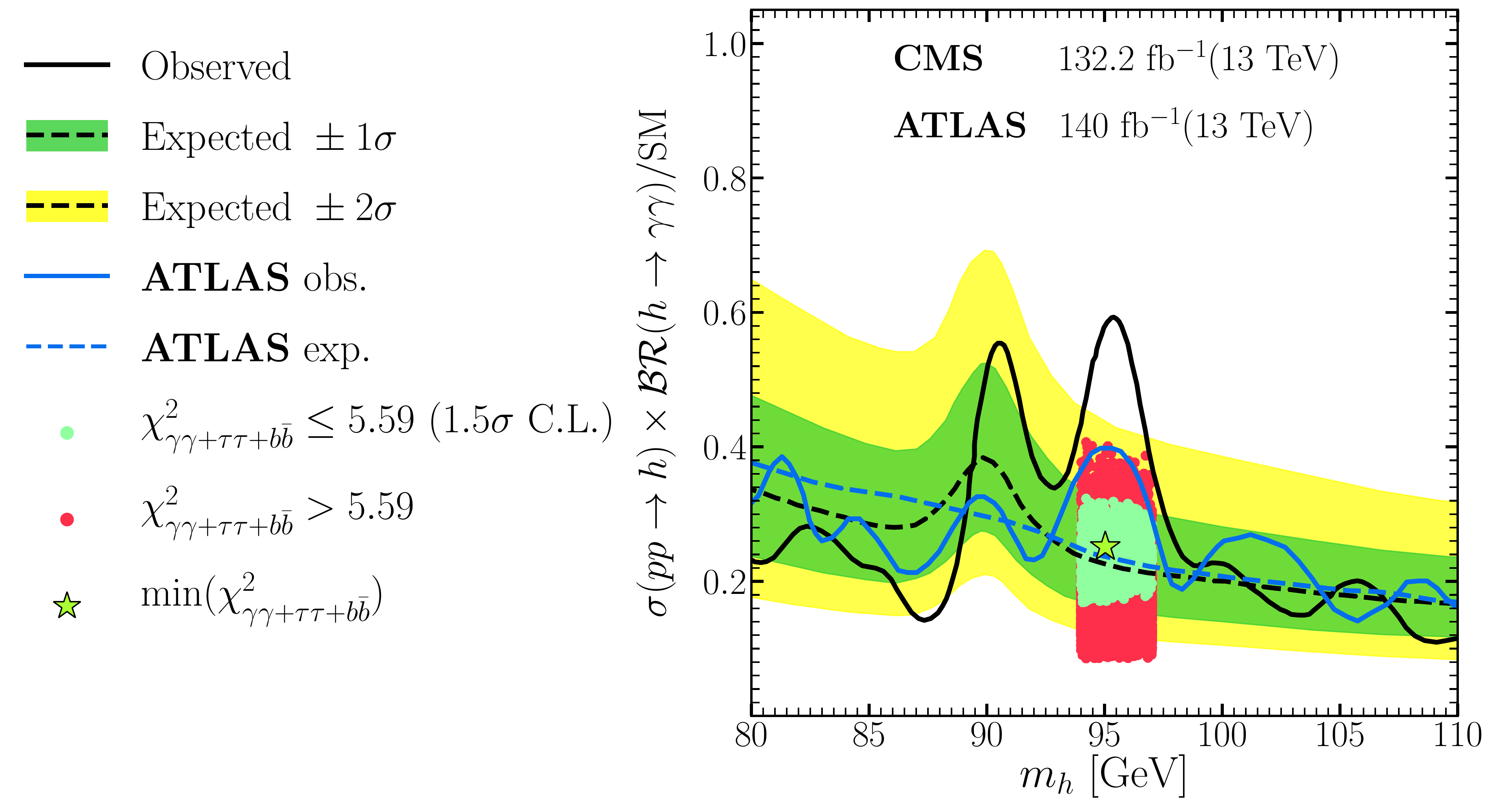}		
		\caption{Allowed points, following the discussed theoretical and experimental constraints,
			superimposed onto the results of the CMS 13 TeV low-mass  $\gamma\gamma$  \cite{CMS:2023yay} analysis. {(Notably, the plot further includes the depiction of the ATLAS expected and observed limits from \cite{Arcangeletti}, showcased in blue.)}
			The light green colour represents the parameter points that fit the excesses within a
			three-dimensional C.L. of 1.5$\sigma$ ($\chi^2_{95}\le 5.59$) {or better, reaching up to $1.3\sigma$},
			whereas the points that fit the excesses at
			2$\sigma$ or more are shown in red. }\label{fig5}
	\end{figure}
	Fig.~\ref{fig6} depicts the correlation between the normalised couplings of the $\approx 125$ GeV Higgs  {using the same color scheme detailed in Fig.~\ref{fig5}}. One can see from the plots that the explanation of the three excesses in  $b\overline{b}$, $\tau \tau$ and $\gamma\gamma$ channels  requires an enhancement of the $h_{125}t\overline{t}$ and $h_{125}b\overline{b}$ couplings deviating by $\sim 11.5\%$ from the SM for $t\bar t$ and  by $\sim 8\%$ for $b\bar b$. In contrast, a decrease can be seen in the Higgs coupling to $\tau\tau$ ($c_{h_{125}\tau\tau}$), deviating from the SM value by about 11\%. Furthermore, the plot includes green dashed lines representing the current $1\sigma$ uncertainties of the normalised couplings $c_{h_{125}i\bar j}$, as measured by CMS \cite{CMS:2022dwd}. Additionally, orange and blue ellipses are depicted, illustrating the projected experimental precision for the normalised couplings at the HL-LHC \cite{Cepeda:2019klc} with an integrated luminosity of 3000 fb$^{-1}$ and the projected precision from a combination of data from the HL-LHC and ILC 500, respectively. One should bear in mind that the center of these experimental projections, for  HL-LHC and ILC 500, corresponds to the SM value represented by black diamonds. Clearly, each point that simultaneously describes the three excesses is situated outside the ellipses corresponding to the HL-LHC and ILC 500. Since the points deviate significantly from the SM predictions, the expected precision of the HL-LHC and ILC 500 experiments would allow us to distinguish between the SM-like properties of $h_{125}$ and the $H$ from the 2HDM Type-III model within the parameter range that aligns with the observed excesses.
	Moreover, one can see  that already at the  HL-LHC one will be able either to discover 
	or rule out the scenario which describes the current excess in three channels under study.
	
	\begin{figure}[H]	
		\centering
		\includegraphics[width=0.95\textwidth]{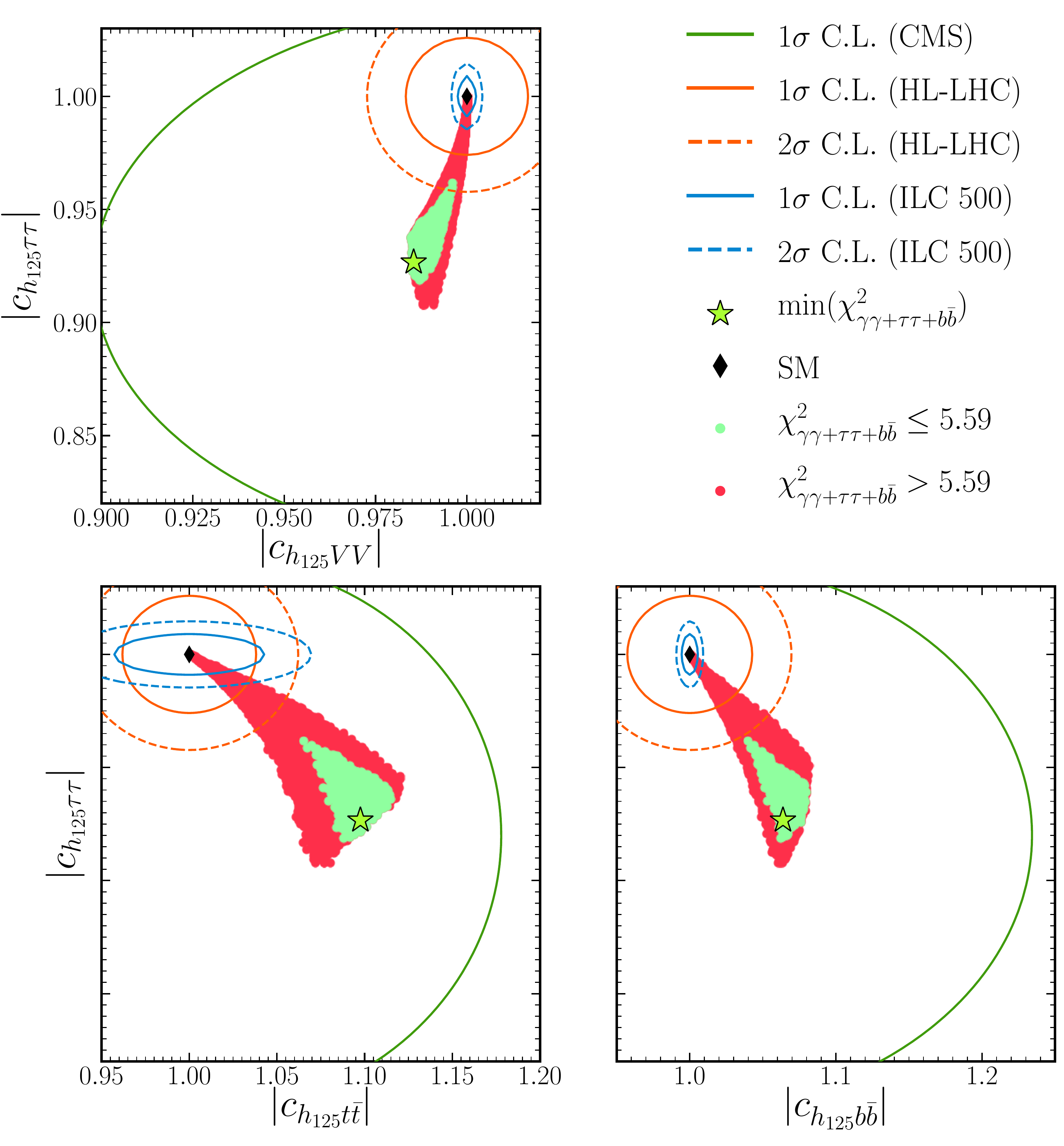}
		\caption{Correlations between the normalized couplings $|c_{h_{125}\tau\tau}|$, $|c_{h_{125}VV}|$, $|c_{h_{125}t\bar{t}}|$, and $|c_{b_{125}t\bar{b}}|$, with colours corresponding to those in Fig.~\ref{fig5}, are illustrated. Also presented are the current 1$\sigma$ uncertainties of the coupling coefficients' measurements from CMS \cite{CMS:2022dwd}, depicted by green dashed lines. The orange and blue ellipses indicate the projected uncertainties at the HL-LHC \cite{Cepeda:2019klc}, and following a combination of data from both the HL-LHC and the ILC at a center-of-mass energy of 500 GeV \cite{Bambade:2019fyw}, respectively.}\label{fig6}
	\end{figure}
	Fig.~\ref{fig7} illustrates the allowed parameter space in the 2HDM Type-III satisfying  theoretical and experimental constraints in the $(m_h,\sigma_{tth_{125}}/\sigma_{tth}^{\mathrm{SM}})$ plane. The black dashed line corresponds to the observed value from CMS \cite{CMS:2022dwd} and the lime green (yellow) band represents instead the $1\sigma$ ($2\sigma$) range. The hatched area denotes the 95\% C.L. probability sensitivity of the HL-LHC \cite{Cepeda:2019klc} to the normalised cross section $\sigma_{tth_{125}}/\sigma_{tth}^{\mathrm{SM}}$, centered on the SM value (solid red line).  It is evident from the figure that all the points that simultaneously explain the three excesses at $1.5\sigma$ C.L {or better}. represented by the light green colour, fall within the 2$\sigma$ measurement range of CMS. Additionally, as seen in Fig.~\ref{fig6}, these points exhibit an enhancement in the $h_{125}{t\bar t}$ coupling. In fact, due to {such an enhancement}, these points deviate significantly from the $2\sigma$ level of the HL-LHC projection. Note, however, that the central value used for the HL-LHC measurement corresponds to the SM prediction, i.e., 1. Thus, for points that {are notably away from the SM prediction}, the projected precision of the HL-LHC experiment would be sufficient to distinguish between the SM-like properties of the $h_{125}$ and the predictions of the 2HDM Type-III within the parameter space consistent with the observed enhancement in the $h_{125}{t\bar t}$ coupling.
	\begin{figure}[H]	
		\centering
		\includegraphics[width=0.95\textwidth]{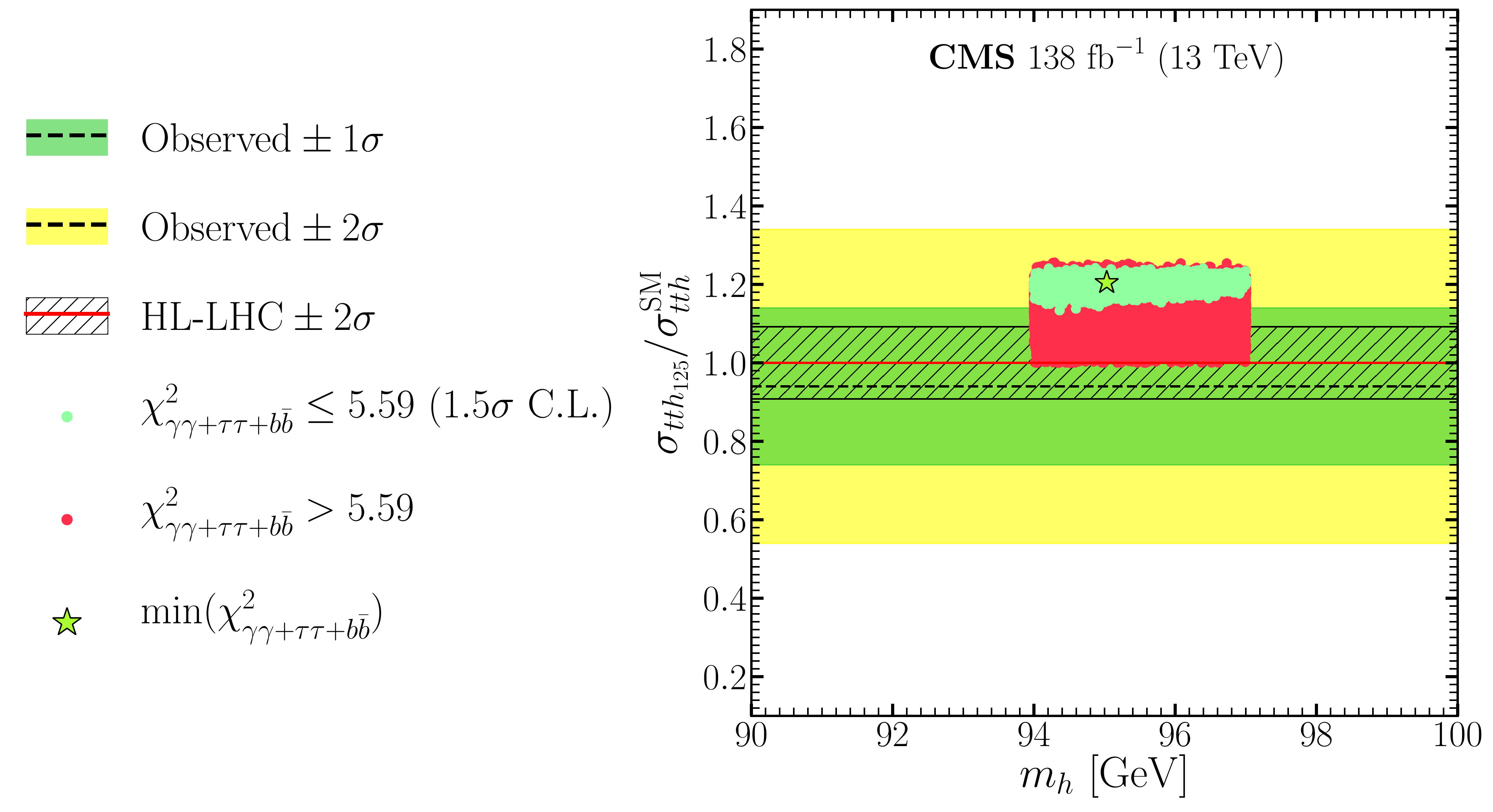}
		\caption{Predicted signal strength $\sigma_{tth_{125}}/\sigma_{tth}^{\mathrm{SM}}$ in relation to the three observed excesses. The black dashed line represents the observed value from CMS \cite{CMS:2022dwd} while the lime green (yellow) band denotes the $1.5\sigma$ ($2\sigma$) range around the observed value. The hatched area illustrates the 95\% probability sensitivity of the HL-LHC \cite{Cepeda:2019klc} to the normalised cross section $\sigma_{tth_{125}}/\sigma_{tth}^{\mathrm{SM}}$, centred on the value of the SM (depicted as a solid red line). The colours correspond to those in Fig.~\ref{fig5}.}\label{fig7}

	\end{figure}

	In summary, there is a smoking-gun prediction stemming from our 2HDM Type-III scenario{, which simultaneously explains $\gamma\gamma, \tau\tau$ and $b\bar b$ anomalies at 1.3 $\sigma$ C.L. and can be tested at the LHC.}
	On the one hand,  $c_{h_{125}t\bar t}$ and $c_{h_{125}b\bar b}$ are consistently larger  than 1, {while} $c_{h_{125}\tau\tau}$ is consistently smaller than 1. Among these couplings, $c_{h_{125}t\bar t}$ can be precisely measured in various production modes}, not only in gluon-gluon fusion where, however, there is contamination from $b$-quark loops, but also in associated production with $t\bar t$ pairs, which could then be tested at the (HL-)LHC, via $pp\to t\bar t h_{125}$, and an ILC 500, via $e^+e^-\to t\bar t h_{125}$.  


Finally, we conclude this section by providing a detailed overview of our best fit point in Tab. \ref{Bp}. One can observe from Tab. \ref{Bp} that the best fit point features a small negative value for the $\chi^u_{33}$ parameter. This, in combination with the small negative $\sin(\beta-\alpha)$ values preferred in our parameter scan gives rise to the observed enhancement in the $h_{125}{t\bar t}$ couplings (see Tab. \ref{coupIII}). 

\section{Conclusions}\label{con}

To date, large data samples have been accumulated by the LHC experiments and many analyses have been performed to examine the discussed 95 GeV excesses, following initial observations. In order to better explain the nature of these potential anomalies, a careful examination of the data, in-depth simulations and advanced computational approaches have been carried out in the literature. Along these lines, we have in this article proposed a theoretical framework, the 2HDM Type-III with a specific Yukawa texture, as a possible solution to the $\gamma\gamma$, $\tau\tau$ and $b\bar b$ anomalies. Specifically, we concentrated on a Higgs boson with a mass of about 95 GeV produced by gluon-gluon fusion at the 13 TeV LHC and decaying into $\tau\tau$ and $\gamma\gamma$ as well as produced by Higgs-strahlung at LEP and decaying into $b\bar b$.

By assuming that the heaviest CP-even Higgs state, $H$, is the one discovered at the LHC with mass $\approx125$ GeV, we have identified parameter space regions where the lightest CP-even state, $h$, with a  mass of $\approx 95$ GeV, {can explain simultaneously the observed excesses through a $\chi^2$ analysis  at approximately 1.3 $\sigma$ C.L. while accommodating both  standard theoretical requirements of self-consistency and  up-to-date experimental constraints.}

Throughout this paper, we have analysed correlations amongst the signal strengths $\mu_{\gamma\gamma}$, $\mu_{\tau\tau}$ and $\mu_{b\bar b}$ at 1$\sigma$ level, arguing that the results presented are compelling and support a more extensive investigation of the proposed 2HDM Type-III scenario, by looking at processes predicted therein, which would constitute an hallmark signature of it, like $pp\to t\bar t H$ production, at both Run 3 of the LHC and the HL-LHC, which would be significantly enhanced with respect to the SM yield. 

In fact, also the study of $e^+e^-\to t\bar tH$ at the ILC 500, which would probe the coupling of $H$ to top (anti)quarks at the percent level, will play a crucial role in confirming the BSM construct pursued in our investigation. Thus, by incorporating the insights gained from these complementary measurements, a more comprehensive understanding of the nature of the studied excesses and the underlying theoretical framework can be achieved (assuming their persistence in future data samples at these machines).

Moreover, the anticipated precision of  the HL-LHC and ILC 500 allows for effective differentiation between the SM-like characteristics of the $H$ state and the predictions of the 2HDM Type-III. This distinction is achievable for data points that display notable deviations from the SM predictions, while remaining within the parameter space that corresponds to the observed enhancement in the ${t\bar t}H$ coupling. Finally, 
{one should stress that the precise measurement of the ${t\bar t} H $ coupling
	would be also  very instrumental for discovery or 
	dismissal of  the proposed scenario already in the near future at the HL-LHC.}

To aid such investigations, we have presented the details of our best fit point for further phenomenological studies in these directions.

\section*{Acknowledgments}
AB and SM are supported in part through the NExT Institute and STFC CG ST/L000296/1. AB would like to thank Prof. Glen Cowan for discussions around the statistical aspects of our study. The work of RB and MB is supported by the Moroccan Ministry of Higher Education and Scientific Research MESRSFC and CNRST Project PPR/2015/6. MB
acknowledges the use of CNRST/HPC-MARWAN in completing this work. SS is supported in full by the NExT Institute.

\definecolor{brilliantrose}{rgb}{1.0, 0.33, 0.64}
\definecolor{lawngreen}{rgb}{0.49, 0.99, 0.0}
\begin{table}[ht]
	\begin{center}
		\setlength{\tabcolsep}{90pt}
		\renewcommand{\arraystretch}{0.1}
		\begin{adjustbox}{max width=0.9\textwidth}		
			\begin{tabular}{lc}
				\hline\hline
				Parameters &        \raisebox{-0.6ex}  {\color{lawngreen}\FiveStar\hspace{-1em}\color{black}\textbf{\FiveStarOpen}}  \\\hline\hline
				\multicolumn{2}{c}{(Masses are in GeV)} \\\hline\hline
				$m_h$   &   95.03  \\
				$m_H$   &   125.09  \\
				$m_A$   &    94.77 \\
				$m_{H^\pm}$   &  162.95  \\
				$\tan\beta$ &   1.74  \\
				$\sin(\beta-\alpha)$  &   -0.17 \\

				$\chi_{11}^u$  & 0.02\\
				$\chi_{22}^u$  &  0.54\\
				$\chi_{33}^u$ & -0.08\\
				$\chi_{11}^d$& -0.41\\
				$\chi_{22}^d$ &   0.24\\
				$\chi_{33}^d$  &  1.55\\
				$\chi_{11}^\ell$ &  -0.06\\
				$\chi_{22}^\ell$ &   0.33\\
				$\chi_{33}^\ell$&    0.97\\

				\hline\hline\multicolumn{2}{c}{Effective coupling $c_{h_{125}i\bar i}$} \\\hline\hline
				$c_{h_{125}t\bar{t}}$ &  1.10   \\
				$c_{h_{125}b\bar{b}}$ &  1.06  \\
				$c_{h_{125}\tau\tau}$ &   0.92   \\

				\hline\hline\multicolumn{2}{c}{Collider signal strength} \\\hline\hline
				$\mu_{\gamma\gamma}$ & 0.25   \\
				$\mu_{\tau\tau}$ & 0.51   \\
				$\mu_{b\bar{b}}$ &  0.02  \\
				\hline\hline\multicolumn{2}{c}{Total decay width in MeV} \\\hline\hline
				$\Gamma(h)$   & 0.27   \\
				$\Gamma(H)$  &4.73   \\
				$\Gamma(A)$  & 0.66    \\
				$\Gamma(H^{\pm})$ & 4.77  \\
				\hline\hline	\multicolumn{2}{c}{${\cal BR}(h\to XY)$ in \%} \\\hline\hline
				${\cal BR}(h\to \gamma\gamma)$   &  0.15  \\
				${\cal BR}(h\to gg)$     & 13.82   \\
				${\cal BR}(h\to b\bar{b})$     &65.37 \\
				${\cal BR}(h\to c\bar{c})$     &    $-$ \\
				${\cal BR}(h\to s\bar{s})$     &   0.50  \\
				${\cal BR}(h\to\mu^+\mu^-)$   &  0.67  \\
				${\cal BR}(h\to\tau\tau)$ & 19.33   \\
				${\cal BR}(h\to ZZ)$      &   $-$   \\
				${\cal BR}(h\to W^+W^-)$     &  0.12    \\
				
				\hline\hline	\multicolumn{2}{c}{${\cal BR}(H\to XY)$ in \%} \\\hline\hline
				${\cal BR}(H\to \gamma\gamma)$   &   0.16   \\
				${\cal BR}(H\to gg)$     &8.16  \\
				${\cal BR}(H\to b\bar{b})$     & 64.08  \\
				${\cal BR}(H\to c\bar{c})$     & 3.07  \\
				${\cal BR}(H\to \tau\tau)$ &  4.70  \\
				${\cal BR}(H\to ZZ)$     &   2.19  \\
				${\cal BR}(H\to W^+W^-)$     &  17.48  \\
				
				\hline\hline	\multicolumn{2}{c}{${\cal BR}(A\to XY)$ in \%} \\\hline\hline
				${\cal BR}(A\to \gamma\gamma)$   &  0.046  \\
				${\cal BR}(A\to gg)$     &  28.70  \\
				${\cal BR}(A\to b\bar{b})$     & 66.95  \\
				${\cal BR}(A\to c\bar{c})$     &   0.40  \\
				${\cal BR}(A\to \mu\mu)$   &   0.16    \\
				${\cal BR}(A\to \tau\tau)$ &   3.53   \\
				
				\hline\hline\multicolumn{2}{c}{${\cal BR}(H^{\pm}\to XY)$ in \%} \\\hline\hline
				${\cal BR}(H^{\pm}\to cs)$   &    0.11  \\
				${\cal BR}(H^{\pm}\to W^+h)$   &   33.79  \\
				${\cal BR}(H^{\pm}\to W^+A)$   & 35.62   \\
				${\cal BR}(H^{\pm}\to\tau\nu)$ &  0.84  \\
				${\cal BR}(H^{\pm}\to tb)$   &  29.47   \\\hline\hline  
				
			\end{tabular}
		\end{adjustbox}
	\end{center}
	\caption{The full description of the best fit point $\chi^2_{95, \mathrm{min}}$ (green star).}\label{Bp}
\end{table}
\bibliography{main} 
\bibliographystyle{JHEP}

\end{document}